\documentclass[10pt,a4paper,twoside]{article}
\usepackage{baltlat5}
\begin{document}
\ \
\vspace{-0.5mm}

\setcounter{page}{293}
\vspace{-2mm}

\titlehead{Baltic Astronomy, vol.\ts 17, 293 (2008).}

\titleb{MULTI-EPOCH UBVRcIc PHOTOMETRIC CATALOG OF SYMBIOTIC STARS}

\begin{authorl}
\authorb{Arne~Henden}{1}
\authorb{and Ulisse~Munari}{2}
\end{authorl}

\begin{addressl}

\addressb{1}{AAVSO, 49 Bay State Road, Cambridge, MA, USA}

\addressb{2}{INAF Osservatorio Astronomico di Padova, via dell'Osservatorio
8, 36012 Asiago (VI), Italy}

\end{addressl}

\submitb{Received 2008 October 13; accepted 2008 November 18}

\begin{summary}

We present a multi-epoch, accurate UBVRcIc photometric catalog of 83
symbiotic stars and related objects, measured while calibrating the Henden
and Munari (2000, 2001, 2006) photometric sequences around them. The vast
majority of the observations where collected in the time interval between
October 19, 1998 to April 21, 2002 with the 1.0-m Ritchey-Chr\'etien
telescope of the U. S. Naval Observatory, Flagstaff Station (Arizona). On
average, UBVRcIc data are given on 3.6 epochs for each program star. The
overall budget error of the data is usually better than 0.01~mag.

\end{summary}


\sectionb{1}{INTRODUCTION}

Symbiotic stars are binary systems composed of a cool giant and a hot
compact companion (generally an accreting white dwarf). They show
variability over any time scale from minutes (flickering) to several decades
(outbursts of symbiotic novae), with phenomena related to the orbital motion
having periodicities generally between 1 and 4 years (or up to a few decades
in the systems harboring a Mira variable, $\sim$20\% of all known symbiotic
stars).

Therefore, the reconstruction of the photometric evolution of any given
symbiotic star, involves combining data from different sources extending
over the widest time interval. Given the generally variable and quite red
colors of symbiotic stars, combining observations from many different
sources would greatly benefit from the use of accurate, well established and
common photometric sequences available around each symbiotic star.

We have established accurate UBVRcIc photometric comparison sequences around
83 mostly northern, symbiotic stars and related objects (Henden and Munari
2000, 2001, 2006). The sequences were intended to assist both current CCD
photometry (therefore covering wide range in colors to firmly anchor the
transformation from local to international photometric systems) as well as
inspection of archival photographic plates (therefore covering, particularly
in the B band, a large range in magnitude).

While establishing the photometric comparison sequences via CCD
observations, we also recorded the corresponding symbiotic stars, and in
this paper we report about their UBVRcIc magnitudes. The observations
presented here were primarily obtained during the period from October 19,
1998 to April 21, 2002.  Some PU Vul observations were obtained in 1995,
some Z And observations occurred during 2003 and 2004, and we have included
more recent observations of RS Oph and Z And.

A previous, multi-epoch photometric catalog of symbiotic stars was published
by Munari et al. (1992). It contained data on 78 mostly southern objects,
observed during 1990 in UBVRcIc from ESO and JHKL from SAAO, and in addition
UBVRj and JHKL data of further 20 northern objects observed from Crimean
Astrophysical Observatory from 1978 to 1989.

The present and Munari et al. (1992) catalogs therefore complements each
other, and together they survey the majority of known symbiotic stars listed
by Belczynski et al. (2000).

\sectionb{2}{OBSERVATIONS}

All of the observations from 2006 forward were acquired using the Sonoita
Research Observatory (SRO) robotic 35cm telescope as part of a calibration
and monitoring campaign.  This telescope has a SBIG STL-1001E CCD camera
with 1.25arcsec pixels and 20x20arcmin field of view.  All prior
observations (the vast majority of Table~1) were made with the
1.0m Ritchey-Chre\'tien telescope of the U.S. Naval Observatory, Flagstaff
Station (NOFS).  Two SITe/Tektronix thinned, backside illuminated CCDs were
used, with 0.6763 arcsec/pixel telescope scale: a 1024x1024 format detector,
yielding 11.4x11.4arcmin field of view, and a 2048x2048 CCD, with a 23x23
arcmin field of view.  Typical seeing was ~2arcsec.  For both SRO and NOFS,
images were processed using IRAF along with nightly median sky flats, dark
and bias frames.  Aperture photometry was performed whenever possible, while
psf-fitting photometry was adopted in the case of blending of the symbiotic
star with nearby field stars (e.g. UV Aur, V4074 Sgr). Inhomogenous
photometry similar to Honeycutt (1992) was used for the final magnitude
determination.  The reported photometry only uses data collected on
photometric nights (transformation errors under 0.02mag).  For each such
night, symbiotic field observations were interspersed with observations of
Landolt (1983,1992) standard fields, selected for wide color and airmass
range.  See Henden and Munari (2000, 2001, 2006) for further details.

\sectionb{3}{THE CATALOG}

The photometric catalog is presented in Table~1, with the program stars
listed in alphabetical order. There, HJD is the heliocentric Julian day
minus 2400000, and refers to the UT middle of the V-band exposure. The
listed errors are the Poissonian component of the total error budget.  
The total error budget is dominated by the transformation from the Landolt
equatorial standard fields, which uncertainty does not in general
exceede 0.02~mag.

The emphasis of Henden and Munari (2000, 2001, 2006) work was to calibrate
the best possible photometric comparison sequences, not to go for the best
measurement of the symbiotic stars. Consequently, for a few stars and for
some of the bands or observing epochs, the symbiotic star may be saturated
or underexposed, and therefore will be missing from Table 1.

The observations collected on V1261~Ori were characterized by the star being
always saturated at VRcIc. It was properly exposed only at U and B bands,
and the corresponding data (not transformed for color equations) are
presented in Table~2.

\begin{center}
\vbox{\footnotesize
\tabcolsep=15pt
\begin{tabular}{@{~~}l@{~~}c@{~~}r@{~~}r@{~~}r@{~~}r@{~~}r@{~~}}
\multicolumn{7}{c}{\parbox{90mm}{
{\normbf \ \ Table 1.}{\norm\ Photometry of the symbiotic stars.}}}\\
\tablerule
\multicolumn{1}{c}{name}&
\multicolumn{1}{c}{HJD}&
\multicolumn{1}{c}{V}&
\multicolumn{1}{c}{B-V}&
\multicolumn{1}{c}{U-B}&
\multicolumn{1}{c}{V-R$_c$}&
\multicolumn{1}{c}{R$_c$-I$_c$}\\
\tablerule
AG Dra               & 52337.9564 &  9.813 ~~0.001 &  1.340 ~~0.002 &  0.319 ~~0.002 &  0.870 ~~0.001 &  0.636 ~~0.003 \\
AG Dra               & 52382.8642 &  9.829 ~~0.001 &  1.376 ~~0.001 &  0.412 ~~0.002 &  0.874 ~~0.001 &  0.670 ~~0.002 \\
AG Dra               & 52383.8746 &  9.826 ~~0.001 &  1.383 ~~0.001 &  0.425 ~~0.003 &  0.860 ~~0.001 &  0.689 ~~0.002 \\
AG Peg               & 52186.7081 &  8.769 ~~0.001 &  1.260 ~~0.001 &--0.012 ~~0.003 &  1.061 ~~0.002 &  1.159 ~~0.003 \\
AG Peg               & 52286.5979 &  8.565 ~~0.002 &  1.210 ~~0.002 &--0.290 ~~0.003 &  1.013 ~~0.003 &  1.085 ~~0.004 \\
AG Peg               & 52288.5591 &  8.588 ~~0.000 &  1.178 ~~0.000 &--0.351 ~~0.002 &  1.047 ~~0.000 &  0.877 ~~0.003 \\
AG Peg               & 52292.5561 &  8.648 ~~0.001 &  1.121 ~~0.001 &--0.343 ~~0.002 &  0.950 ~~0.002 &  0.970 ~~0.003 \\
ALS 1                & 51456.6970 & 13.518 ~~0.002 &  1.381 ~~0.003 &  0.329 ~~0.008 &  1.156 ~~0.002 &  1.373 ~~0.002 \\
ALS 1                & 51457.6891 & 13.516 ~~0.002 &  1.376 ~~0.003 &  0.318 ~~0.005 &  1.158 ~~0.001 &  1.377 ~~0.001 \\
ALS 1                & 51464.6866 & 13.428 ~~0.002 &  1.409 ~~0.003 &  0.295 ~~0.005 &  1.147 ~~0.003 &  1.354 ~~0.001 \\
ALS 2                & 51455.6396 & 14.324 ~~0.004 &  1.899 ~~0.007 &  0.529 ~~0.025 &  1.374 ~~0.004 &  1.211 ~~0.005 \\
ALS 2                & 51456.5950 & 14.335 ~~0.004 &  1.890 ~~0.007 &  0.469 ~~0.014 &  1.373 ~~0.005 &  1.227 ~~0.003 \\
ALS 2                & 51459.5961 & 14.302 ~~0.005 &  1.882 ~~0.011 &  0.546 ~~0.024 &  1.346 ~~0.004 &  1.229 ~~0.004 \\
ALS 2                & 51465.5988 & 14.315 ~~0.005 &  1.905 ~~0.010 &  0.428 ~~0.025 &  1.393 ~~0.005 &  1.228 ~~0.005 \\
ALS 2                & 51486.5555 & 14.332 ~~0.008 &  1.913 ~~0.023 &  0.523 ~~0.069 &  1.370 ~~0.008 &  1.269 ~~0.009 \\
Ap 3-1               & 51455.7309 & 17.008 ~~0.036 &  2.026 ~~0.084 &  0.929 ~~0.155 &  2.153 ~~0.038 &  2.178 ~~0.013 \\
Ap 3-1               & 51456.7230 & 17.024 ~~0.044 &  2.125 ~~0.105 &  0.572 ~~0.158 &  2.139 ~~0.046 &  2.200 ~~0.015 \\
Ap 3-1               & 51457.7087 & 17.026 ~~0.023 &  2.072 ~~0.062 &  0.685 ~~0.099 &  2.125 ~~0.024 &  2.243 ~~0.009 \\
AS 201               & 51999.6488 & 11.834 ~~0.001 &  0.702 ~~0.001 &  0.113 ~~0.003 &  0.387 ~~0.001 &  0.337 ~~0.003 \\
AS 201               & 52014.6273 & 11.835 ~~0.001 &  0.690 ~~0.002 &  0.117 ~~0.003 &  0.385 ~~0.001 &  0.327 ~~0.003 \\
AS 201               & 52016.6320 & 11.837 ~~0.001 &  0.696 ~~0.001 &  0.115 ~~0.003 &  0.381 ~~0.001 &  0.332 ~~0.002 \\
AS 210               & 52052.8452 & 12.123 ~~0.003 &  1.482 ~~0.004 &--0.478 ~~0.004 &  1.384 ~~0.002 &  0.929 ~~0.002 \\
AS 210               & 52061.8452 & 12.225 ~~0.002 &  1.409 ~~0.004 &--0.520 ~~0.005 &  1.421 ~~0.001 &  0.942 ~~0.001 \\
AS 210               & 52115.6991 & 12.617 ~~0.003 &  1.127 ~~0.006 &--0.572 ~~0.004 &  1.467 ~~0.003 &  0.982 ~~0.002 \\
AS 289               & 52153.7092 & 12.700 ~~0.002 &  1.621 ~~0.004 &--0.021 ~~0.006 &  1.571 ~~0.003 &  1.721 ~~0.001 \\
AS 289               & 52157.6347 & 12.687 ~~0.002 &  1.608 ~~0.004 &--0.021 ~~0.004 &  1.576 ~~0.002 &  1.714 ~~0.001 \\
AS 289               & 52171.6437 & 12.691 ~~0.002 &  1.612 ~~0.003 &--0.012 ~~0.004 &  1.553 ~~0.003 &  1.745 ~~0.001 \\
AS 323               & 51456.6747 & 14.215 ~~0.006 &  0.964 ~~0.009 &--0.392 ~~0.007 &  1.097 ~~0.007 &  1.375 ~~0.005 \\
AS 323               & 51457.6401 & 14.212 ~~0.005 &  0.994 ~~0.006 &--0.492 ~~0.007 &  1.120 ~~0.005 &  1.397 ~~0.007 \\
AS 323               & 51465.6534 & 14.145 ~~0.006 &  1.013 ~~0.007 &--0.380 ~~0.009 &  1.100 ~~0.006 &  1.389 ~~0.005 \\
AS 327               & 52157.6667 & 12.861 ~~0.002 &  1.571 ~~0.005 &  0.244 ~~0.008 &  1.265 ~~0.004 &  1.115 ~~0.003 \\
AS 327               & 52171.5998 & 13.029 ~~0.003 &  1.470 ~~0.007 &  0.076 ~~0.006 &  1.282 ~~0.005 &  1.173 ~~0.004 \\
AS 327               & 52174.6254 & 13.031 ~~0.003 &  1.471 ~~0.005 &  0.052 ~~0.007 &  1.283 ~~0.004 &  1.162 ~~0.003 \\
AX Per               & 51873.6822 & 11.341 ~~0.002 &  1.271 ~~0.002 &--0.380 ~~0.004 &  1.321 ~~0.002 &  1.520 ~~0.001 \\
AX Per               & 51877.6873 & 11.340 ~~0.002 &  1.272 ~~0.003 &--0.383 ~~0.004 &  1.310 ~~0.004 &  1.331 ~~0.005 \\
AX Per               & 51906.6660 & 11.440 ~~0.002 &  1.224 ~~0.003 &--0.504 ~~0.004 &  1.320 ~~0.002 &  1.547 ~~0.001 \\
BD Cam               & 52287.7223 &  8.732 ~~0.001 &  1.138 ~~0.002 &                &  0.661 ~~0.002 &                \\
BF Cyg               & 52110.9205 & 12.441 ~~0.002 &  0.804 ~~0.004 &--0.506 ~~0.003 &  1.181 ~~0.002 &  1.504 ~~0.002 \\
BF Cyg               & 52145.8160 & 12.595 ~~0.002 &  0.747 ~~0.003 &--0.512 ~~0.004 &  1.157 ~~0.002 &  1.528 ~~0.001 \\
BF Cyg               & 52146.8206 & 12.586 ~~0.002 &  0.729 ~~0.003 &--0.506 ~~0.003 &  1.153 ~~0.002 &  1.527 ~~0.002 \\
BF Cyg               & 52153.7988 & 12.648 ~~0.002 &  0.716 ~~0.003 &--0.485 ~~0.003 &  1.174 ~~0.004 &  1.484 ~~0.003 \\
BX Mon               & 51581.7547 & 11.281 ~~0.001 &  1.003 ~~0.001 &--0.441 ~~0.001 &  1.057 ~~0.002 &  1.489 ~~0.002 \\
BX Mon               & 51614.7060 & 11.235 ~~0.001 &  1.162 ~~0.001 &--0.318 ~~0.002 &  1.133 ~~0.002 &  1.454 ~~0.002 \\
BX Mon               & 51629.7021 & 11.176 ~~0.001 &  1.216 ~~0.002 &--0.351 ~~0.002 &  1.150 ~~0.001 &  1.452 ~~0.002 \\
\tablerule
\end{tabular}
}
\end{center}
\vspace{-2mm}
\begin{center}
\vbox{\footnotesize
\tabcolsep=15pt
\begin{tabular}{@{~~}l@{~~}c@{~~}r@{~~}r@{~~}r@{~~}r@{~~}r@{~~}}
\multicolumn{7}{c}{\parbox{90mm}{
{\normbf \ \ Table 1.}{\norm\ (continues).}}}\\
\tablerule
\multicolumn{1}{c}{name}&
\multicolumn{1}{c}{HJD}&
\multicolumn{1}{c}{V}&
\multicolumn{1}{c}{B-V}&
\multicolumn{1}{c}{U-B}&
\multicolumn{1}{c}{V-R$_c$}&
\multicolumn{1}{c}{R$_c$-I$_c$}\\
\tablerule
CH Cyg               & 52382.9451 &  7.560 ~~0.003 &  1.679 ~~0.002 &  0.878 ~~0.002 &  1.766 ~~0.002 &                \\
CH Cyg               & 52383.9103 &  7.565 ~~0.002 &  1.688 ~~0.001 &  0.869 ~~0.002 &  1.694 ~~0.002 &                \\
CH Cyg               & 52385.9986 &  7.650 ~~0.002 &  1.660 ~~0.001 &  0.869 ~~0.001 &  1.643 ~~0.001 &                \\
CI Cam               & 51220.7178 & 11.602 ~~0.002 &  0.820 ~~0.002 &                &                &                \\
CI Cam               & 51220.7203 & 11.604 ~~0.002 &  0.808 ~~0.001 &                &                &                \\
CI Cam               & 51222.7364 & 11.616 ~~0.002 &  0.803 ~~0.001 &                &                &                \\
CI Cam               & 51877.7400 & 11.725 ~~0.001 &  0.793 ~~0.002 &--0.336 ~~0.002 &  0.887 ~~0.001 &                \\
CI Cam               & 51906.6750 & 11.718 ~~0.002 &  0.809 ~~0.002 &--0.366 ~~0.003 &  0.892 ~~0.001 &  0.672 ~~0.002 \\
CI Cam               & 51930.6738 & 11.715 ~~0.001 &  0.822 ~~0.001 &--0.284 ~~0.004 &  0.889 ~~0.002 &  0.693 ~~0.001 \\
CI Cyg               & 52110.9312 & 11.096 ~~0.002 &  1.539 ~~0.003 &  0.111 ~~0.004 &  1.471 ~~0.002 &  1.738 ~~0.002 \\
CI Cyg               & 52145.8798 & 11.240 ~~0.001 &  1.390 ~~0.002 &--0.086 ~~0.003 &  1.484 ~~0.001 &  1.834 ~~0.002 \\
CI Cyg               & 52146.8590 & 11.256 ~~0.001 &  1.390 ~~0.002 &--0.102 ~~0.003 &  1.481 ~~0.001 &  1.843 ~~0.002 \\
CM Aql               & 51455.7010 & 13.385 ~~0.001 &  1.275 ~~0.003 &--0.260 ~~0.005 &  1.148 ~~0.001 &  1.207 ~~0.003 \\
CM Aql               & 51457.6682 & 13.347 ~~0.002 &  1.262 ~~0.003 &--0.280 ~~0.004 &  1.136 ~~0.002 &  1.212 ~~0.001 \\
CM Aql               & 51464.6664 & 13.371 ~~0.002 &  1.281 ~~0.003 &--0.301 ~~0.003 &  1.146 ~~0.002 &  1.222 ~~0.002 \\
DQ Ser               & 51729.7685 & 14.966 ~~0.004 &  1.334 ~~0.007 &  0.319 ~~0.010 &  1.296 ~~0.005 &  1.668 ~~0.004 \\
DQ Ser               & 51730.7943 & 14.943 ~~0.003 &  1.343 ~~0.006 &  0.291 ~~0.010 &  1.305 ~~0.004 &  1.630 ~~0.004 \\
DQ Ser               & 51731.7670 & 14.909 ~~0.003 &  1.348 ~~0.006 &  0.304 ~~0.009 &  1.301 ~~0.004 &  1.626 ~~0.004 \\
DT Ser               & 51728.8404 & 16.214 ~~0.047 &  0.378 ~~0.056 &--0.832 ~~0.033 &--0.062 ~~0.057 &  0.094 ~~0.049 \\
DT Ser               & 51729.8376 & 16.216 ~~0.052 &  0.375 ~~0.059 &--0.821 ~~0.032 &--0.035 ~~0.058 &                \\
DT Ser               & 51730.7814 & 16.210 ~~0.047 &  0.361 ~~0.053 &--0.807 ~~0.031 &--0.067 ~~0.056 &--0.073 ~~0.045 \\
Draco C-1            & 51315.8097 & 17.118 ~~0.006 &  1.450 ~~0.013 &--0.013 ~~0.024 &  0.886 ~~0.008 &  0.567 ~~0.009 \\
Draco C-1            & 51316.8076 & 17.096 ~~0.005 &  1.473 ~~0.010 &--0.003 ~~0.022 &  0.865 ~~0.006 &  0.570 ~~0.006 \\
Draco C-1            & 51317.8164 & 17.081 ~~0.005 &  1.496 ~~0.010 &  0.028 ~~0.025 &  0.830 ~~0.006 &  0.589 ~~0.006 \\
EG And               & 52186.8091 &  7.241 ~~0.002 &  1.605 ~~0.002 &  1.676 ~~0.005 &  0.838 ~~0.003 &                \\
EG And               & 52287.7022 &  7.502 ~~0.001 &  1.525 ~~0.001 &  1.444 ~~0.004 &                &                \\
EG And               & 52288.6575 &  7.463 ~~0.000 &  1.474 ~~0.002 &  1.518 ~~0.006 &                &                \\
ER Del               & 51729.9051 & 10.354 ~~0.001 &  1.765 ~~0.002 &  1.170 ~~0.003 &  1.173 ~~0.002 &  1.478 ~~0.002 \\
ER Del               & 51730.9447 & 10.328 ~~0.001 &  1.785 ~~0.001 &  1.318 ~~0.002 &  1.177 ~~0.002 &  1.454 ~~0.002 \\
ER Del               & 51731.8077 & 10.308 ~~0.001 &  1.797 ~~0.002 &  1.356 ~~0.003 &  1.170 ~~0.002 &  1.453 ~~0.002 \\
FG Ser               & 51247.9919 & 11.872 ~~0.001 &  1.683 ~~0.002 &--0.119 ~~0.014 &  1.691 ~~0.001 &  1.626 ~~0.003 \\
FG Ser               & 51251.9924 & 11.886 ~~0.002 &  1.673 ~~0.002 &--0.169 ~~0.003 &  1.597 ~~0.002 &  1.554 ~~0.001 \\
FG Ser               & 51403.7917 & 11.633 ~~0.001 &  1.755 ~~0.002 &--0.082 ~~0.010 &  1.686 ~~0.001 &  1.712 ~~0.001 \\
FG Ser               & 51465.6639 & 11.700 ~~0.001 &  1.785 ~~0.003 &  0.071 ~~0.005 &  1.709 ~~0.001 &  1.617 ~~0.002 \\
FG Ser               & 51487.5820 & 11.774 ~~0.001 &  1.804 ~~0.002 &  0.140 ~~0.004 &  1.362 ~~0.003 &  1.517 ~~0.001 \\
FN Sgr               & 51456.6244 & 12.075 ~~0.002 &  0.717 ~~0.002 &--0.115 ~~0.002 &  0.809 ~~0.002 &  0.898 ~~0.001 \\
FN Sgr               & 51457.5924 & 12.120 ~~0.002 &  0.737 ~~0.003 &--0.152 ~~0.003 &  0.825 ~~0.002 &  0.919 ~~0.001 \\
FN Sgr               & 51465.6147 & 12.116 ~~0.004 &  0.731 ~~0.001 &--0.108 ~~0.003 &  0.853 ~~0.002 &  0.903 ~~0.002 \\
FN Sgr               & 51487.5651 & 12.005 ~~0.002 &  0.713 ~~0.002 &--0.107 ~~0.002 &  0.801 ~~0.002 &  0.885 ~~0.001 \\
GH Gem               & 51581.7214 & 13.752 ~~0.002 &  1.042 ~~0.003 &  0.586 ~~0.005 &  0.628 ~~0.003 &  0.621 ~~0.003 \\
GH Gem               & 51613.8025 & 14.159 ~~0.003 &  0.975 ~~0.004 &  0.495 ~~0.008 &  0.575 ~~0.004 &  0.538 ~~0.004 \\
GH Gem               & 51614.7269 & 14.125 ~~0.002 &  0.982 ~~0.004 &  0.516 ~~0.007 &  0.582 ~~0.003 &  0.548 ~~0.004 \\
Hen 3-1341           & 51664.9078 & 11.325 ~~0.003 &  0.506 ~~0.001 &--0.841 ~~0.003 &  0.665 ~~0.002 &  0.616 ~~0.002 \\
Hen 3-1341           & 51666.9428 & 11.361 ~~0.001 &  0.557 ~~0.001 &--0.901 ~~0.002 &  0.628 ~~0.002 &  0.662 ~~0.002 \\
Hen 3-1341           & 51692.8582 & 11.376 ~~0.001 &  0.523 ~~0.001 &--0.797 ~~0.002 &  0.679 ~~0.002 &  0.620 ~~0.002 \\
Hen 2-468            & 51402.8911 & 14.783 ~~0.002 &  1.830 ~~0.004 &  0.614 ~~0.011 &  1.527 ~~0.003 &  1.707 ~~0.002 \\
Hen 2-468            & 51403.8338 & 14.782 ~~0.002 &  1.828 ~~0.004 &  0.603 ~~0.008 &  1.527 ~~0.003 &  1.714 ~~0.002 \\ 
Hen 2-442            & 52145.8433 & 16.050 ~~0.008 &  1.412 ~~0.013 &  0.260 ~~0.018 &  1.085 ~~0.012 &  0.621 ~~0.016 \\
Hen 2-442            & 52171.7666 & 16.062 ~~0.007 &  1.404 ~~0.012 &  0.257 ~~0.018 &  1.088 ~~0.009 &  0.618 ~~0.011 \\
Hen 2-442            & 52174.7610 & 16.052 ~~0.007 &  1.415 ~~0.012 &  0.249 ~~0.020 &  1.096 ~~0.009 &  0.603 ~~0.011 \\
Hen 3-1342           & 52052.8568 & 13.422 ~~0.003 &  1.821 ~~0.005 &  0.278 ~~0.008 &  1.192 ~~0.004 &  1.046 ~~0.003 \\
Hen 3-1342           & 52061.8586 & 13.426 ~~0.003 &  1.822 ~~0.006 &  0.327 ~~0.010 &  1.207 ~~0.004 &  1.056 ~~0.003 \\
Hen 3-1342           & 52115.7371 & 13.418 ~~0.004 &  1.876 ~~0.005 &  0.776 ~~0.009 &  1.186 ~~0.004 &  1.107 ~~0.003 \\
\tablerule
\end{tabular}
}
\end{center}
\vspace{-2mm}
\begin{center}
\vbox{\footnotesize
\tabcolsep=15pt
\begin{tabular}{@{~~}l@{~~}c@{~~}r@{~~}r@{~~}r@{~~}r@{~~}r@{~~}}
\multicolumn{7}{c}{\parbox{90mm}{
{\normbf \ \ Table 1.}{\norm\ (continues).}}}\\
\tablerule
\multicolumn{1}{c}{name}&
\multicolumn{1}{c}{HJD}&
\multicolumn{1}{c}{V}&
\multicolumn{1}{c}{B-V}&
\multicolumn{1}{c}{U-B}&
\multicolumn{1}{c}{V-R$_c$}&
\multicolumn{1}{c}{R$_c$-I$_c$}\\
\tablerule
Hen 3-1591           & 52382.9707 & 13.186 ~~0.004 &  1.217 ~~0.004 &  0.598 ~~0.007 &  0.747 ~~0.004 &  0.704 ~~0.005 \\
Hen 3-1591           & 52383.9378 & 13.157 ~~0.002 &  1.222 ~~0.004 &  0.611 ~~0.006 &  0.751 ~~0.003 &  0.694 ~~0.004 \\
Hen 3-1591           & 52385.9625 & 13.075 ~~0.002 &  1.217 ~~0.004 &  0.587 ~~0.005 &  0.728 ~~0.003 &  0.695 ~~0.004 \\
HM Sge               & 52145.8347 & 11.749 ~~0.002 &  0.812 ~~0.003 &--0.708 ~~0.003 &  0.760 ~~0.003 &--0.122 ~~0.004 \\
HM Sge               & 52146.8303 & 11.756 ~~0.002 &  0.810 ~~0.003 &--0.701 ~~0.003 &  0.761 ~~0.003 &--0.128 ~~0.004 \\
HM Sge               & 52153.8074 & 11.746 ~~0.002 &  0.816 ~~0.003 &--0.685 ~~0.003 &  0.761 ~~0.003 &--0.142 ~~0.004 \\
IV Vir               & 51614.9327 & 10.757 ~~0.001 &  1.440 ~~0.002 &  0.673 ~~0.005 &  0.830 ~~0.001 &  0.732 ~~0.002 \\
IV Vir               & 51629.8606 & 10.687 ~~0.001 &  1.399 ~~0.002 &  0.544 ~~0.003 &  0.809 ~~0.002 &  0.737 ~~0.002 \\
IV Vir               & 51639.8596 & 10.681 ~~0.001 &  1.390 ~~0.002 &  0.403 ~~0.003 &  0.809 ~~0.002 &  0.724 ~~0.001 \\
K 3-9                & 51403.7401 & 16.971 ~~0.007 &  1.289 ~~0.012 &  0.552 ~~0.031 &  0.838 ~~0.011 &  0.961 ~~0.017 \\
K 3-9                & 51455.6808 & 16.914 ~~0.013 &  1.313 ~~0.022 &  0.482 ~~0.051 &  0.880 ~~0.020 &  1.000 ~~0.027 \\
K 3-9                & 51464.6390 & 16.980 ~~0.009 &  1.288 ~~0.015 &  0.425 ~~0.044 &  0.853 ~~0.014 &  0.947 ~~0.021 \\
LL Cas               & 51549.5720 & 15.968 ~~0.013 &  0.886 ~~0.018 &--0.635 ~~0.016 &  1.680 ~~0.015 &  2.257 ~~0.008 \\
LL Cas               & 51550.5736 & 16.045 ~~0.015 &  0.893 ~~0.021 &--0.663 ~~0.019 &  1.737 ~~0.017 &  2.248 ~~0.008 \\
LL Cas               & 51554.6255 & 16.011 ~~0.010 &  0.902 ~~0.014 &--0.644 ~~0.013 &  1.662 ~~0.012 &  2.101 ~~0.007 \\
LT Del               & 51453.7393 & 13.064 ~~0.001 &  1.264 ~~0.002 &--0.231 ~~0.005 &  0.811 ~~0.002 &  0.680 ~~0.003 \\
LT Del               & 51455.7588 & 13.057 ~~0.002 &  1.270 ~~0.003 &--0.216 ~~0.004 &  0.817 ~~0.003 &  0.666 ~~0.003 \\
LT Del               & 51456.7591 & 13.061 ~~0.001 &  1.271 ~~0.002 &--0.208 ~~0.003 &  0.819 ~~0.001 &  0.675 ~~0.002 \\
LW Cas               & 51554.7158 & 17.480 ~~0.014 &  1.802 ~~0.046 &                &  1.202 ~~0.018 &  1.195 ~~0.015 \\
LW Cas               & 51577.7121 & 17.503 ~~0.015 &  1.806 ~~0.051 &                &  1.179 ~~0.019 &  1.283 ~~0.015 \\
LW Cas               & 51874.6722 & 17.390 ~~0.011 &  1.936 ~~0.030 &                &  1.082 ~~0.016 &  1.231 ~~0.014 \\
M 1-21               & 52052.9057 & 14.470 ~~0.004 &  1.074 ~~0.006 &--0.316 ~~0.006 &  1.671 ~~0.005 &  1.469 ~~0.004 \\
M 1-21               & 52115.7581 & 14.714 ~~0.005 &  1.093 ~~0.007 &--0.153 ~~0.008 &  1.665 ~~0.006 &  1.679 ~~0.004 \\
M 1-21               & 52145.6308 & 14.829 ~~0.007 &  1.070 ~~0.009 &--0.181 ~~0.010 &  1.671 ~~0.007 &  1.672 ~~0.004 \\
MaC 1-17             & 52145.7650 & 15.182 ~~0.005 &  1.624 ~~0.010 &--0.034 ~~0.013 &  1.402 ~~0.007 &  1.507 ~~0.007 \\
MaC 1-17             & 52157.7156 & 15.233 ~~0.007 &  1.575 ~~0.015 &--0.123 ~~0.021 &  1.441 ~~0.008 &  1.460 ~~0.006 \\
MaC 1-17             & 52171.7069 & 15.039 ~~0.005 &  1.645 ~~0.009 &--0.131 ~~0.012 &  1.373 ~~0.007 &  1.386 ~~0.006 \\
MWC 960              & 51105.5974 & 12.110 ~~0.002 &  1.626 ~~0.003 &                &                &                \\
MWC 960              & 51106.5818 & 12.122 ~~0.001 &  1.623 ~~0.002 &                &                &                \\
MWC 960              & 51349.8982 & 12.196 ~~0.002 &  1.639 ~~0.003 &  0.261 ~~0.005 &  1.080 ~~0.001 &  0.775 ~~0.002 \\
MWC 960              & 51352.8427 & 12.204 ~~0.002 &  1.640 ~~0.003 &  0.248 ~~0.005 &  1.108 ~~0.002 &  0.762 ~~0.002 \\
MWC 960              & 51465.5739 & 12.197 ~~0.004 &  1.594 ~~0.004 &  0.156 ~~0.008 &  1.088 ~~0.003 &  0.743 ~~0.003 \\
MWC 960              & 51486.5748 & 12.100 ~~0.003 &  1.606 ~~0.003 &  0.285 ~~0.005 &  1.047 ~~0.003 &  0.729 ~~0.004 \\
NQ Gem               & 52287.7964 &  8.133 ~~0.000 &  1.900 ~~0.002 &  2.529 ~~0.003 &  0.974 ~~0.002 &  0.636 ~~0.000 \\
NQ Gem               & 52288.7872 &  7.987 ~~0.000 &  2.015 ~~0.001 &  2.522 ~~0.003 &  0.846 ~~0.000 &  0.779 ~~0.002 \\
NQ Gem               & 52292.7095 &  7.935 ~~0.001 &  2.045 ~~0.002 &  2.539 ~~0.004 &  0.865 ~~0.000 &  0.985 ~~0.002 \\
NQ Gem               & 52292.7791 &  7.892 ~~0.000 &  2.077 ~~0.002 &  2.540 ~~0.003 &  0.900 ~~0.002 &  0.728 ~~0.001 \\
NSV 11776            & 51728.8702 & 13.473 ~~0.003 &  0.921 ~~0.003 &--0.606 ~~0.003 &  1.394 ~~0.002 &  0.602 ~~0.002 \\
NSV 11776            & 51729.8508 & 13.473 ~~0.003 &  0.920 ~~0.003 &--0.602 ~~0.003 &  1.378 ~~0.002 &  0.612 ~~0.002 \\
NSV 11776            & 51730.8298 & 13.472 ~~0.002 &  0.917 ~~0.003 &--0.582 ~~0.004 &  1.394 ~~0.002 &  0.618 ~~0.002 \\
Pe 2-16              & 52157.6930 & 15.931 ~~0.010 &  1.765 ~~0.026 &  0.308 ~~0.051 &  1.705 ~~0.012 &  1.644 ~~0.007 \\
Pe 2-16              & 52171.6764 & 15.844 ~~0.007 &  1.759 ~~0.012 &  0.480 ~~0.021 &  1.672 ~~0.009 &  1.727 ~~0.006 \\
Pe 2-16              & 52174.6594 & 15.838 ~~0.008 &  1.751 ~~0.014 &  0.506 ~~0.024 &  1.685 ~~0.009 &  1.735 ~~0.006 \\
Pt 1                 & 52145.6510 & 15.562 ~~0.011 &  2.080 ~~0.025 &  0.909 ~~0.076 &  1.605 ~~0.013 &  1.591 ~~0.009 \\
Pt 1                 & 52146.6566 & 15.550 ~~0.010 &  2.134 ~~0.026 &  0.804 ~~0.085 &  1.576 ~~0.012 &  1.611 ~~0.008 \\
Pt 1                 & 52153.6477 & 15.603 ~~0.015 &  2.122 ~~0.051 &  0.777 ~~0.193 &  1.667 ~~0.017 &  1.550 ~~0.009 \\
PU Vul               & 49990.7933 & 11.656 ~~0.001 &  0.155 ~~0.002 &                &                &                \\
PU Vul               & 49991.7948 & 11.657 ~~0.002 &  0.143 ~~0.002 &                &                &                \\
PU Vul               & 49992.6915 & 11.674 ~~0.002 &  0.134 ~~0.002 &                &                &                \\
PU Vul               & 51873.5997 & 11.749 ~~0.001 &  0.323 ~~0.001 &--0.812 ~~0.001 &  1.388 ~~0.002 &  0.555 ~~0.001 \\
PU Vul               & 51877.6091 & 11.757 ~~0.001 &  0.319 ~~0.001 &--0.800 ~~0.001 &  1.357 ~~0.001 &  0.601 ~~0.002 \\
PU Vul               & 51877.6186 & 11.764 ~~0.005 &  0.306 ~~0.007 &                &  1.386 ~~0.006 &  0.577 ~~0.004 \\
PU Vul               & 52145.8961 & 11.829 ~~0.001 &  0.388 ~~0.002 &--0.774 ~~0.001 &  1.225 ~~0.002 &  0.934 ~~0.002 \\
\tablerule
\end{tabular}
}
\end{center}
\vspace{-2mm}
\begin{center}
\vbox{\footnotesize
\tabcolsep=15pt
\begin{tabular}{@{~~}l@{~~}c@{~~}r@{~~}r@{~~}r@{~~}r@{~~}r@{~~}}
\multicolumn{7}{c}{\parbox{90mm}{
{\normbf \ \ Table 1.}{\norm\ (continues).}}}\\
\tablerule
\multicolumn{1}{c}{name}&
\multicolumn{1}{c}{HJD}&
\multicolumn{1}{c}{V}&
\multicolumn{1}{c}{B-V}&
\multicolumn{1}{c}{U-B}&
\multicolumn{1}{c}{V-R$_c$}&
\multicolumn{1}{c}{R$_c$-I$_c$}\\
\tablerule
QW Sge               & 51453.7213 & 12.572 ~~0.004 &  0.972 ~~0.005 &--0.193 ~~0.004 &  0.928 ~~0.004 &  1.201 ~~0.005 \\
QW Sge               & 51455.7419 & 12.301 ~~0.003 &  0.758 ~~0.004 &  0.013 ~~0.004 &  0.813 ~~0.004 &  1.089 ~~0.004 \\
QW Sge               & 51456.7499 & 12.262 ~~0.002 &  0.696 ~~0.003 &  0.065 ~~0.003 &  0.787 ~~0.003 &  1.097 ~~0.002 \\
RS Oph               & 52382.9352 & 11.540 ~~0.002 &  1.443 ~~0.005 &                &  1.116 ~~0.002 &  1.037 ~~0.002 \\
RS Oph               & 52383.9041 & 11.375 ~~0.002 &  1.261 ~~0.003 &--0.073 ~~0.003 &  1.030 ~~0.002 &  1.032 ~~0.001 \\
RS Oph               & 52385.9522 & 11.238 ~~0.002 &  1.223 ~~0.003 &  0.059 ~~0.003 &  0.990 ~~0.002 &  0.998 ~~0.002 \\
RS Oph               & 53993.6153 & 11.628 ~~0.003 &  1.513 ~~0.005 &                &  1.065 ~~0.004 &  0.657 ~~0.002 \\
RS Oph               & 53993.6218 & 11.626 ~~0.003 &  1.527 ~~0.005 &                &  1.186 ~~0.005 &  0.609 ~~0.002 \\
RS Oph               & 53995.6550 & 11.485 ~~0.003 &  1.596 ~~0.005 &                &  1.237 ~~0.005 &  0.959 ~~0.005 \\
RS Oph               & 53995.6628 & 11.477 ~~0.004 &  1.594 ~~0.006 &                &  1.213 ~~0.008 &  0.992 ~~0.005 \\
RS Oph               & 53996.6114 & 11.406 ~~0.002 &  1.615 ~~0.009 &                &  1.193 ~~0.005 &  0.959 ~~0.006 \\
RS Oph               & 53996.6193 & 11.415 ~~0.003 &  1.595 ~~0.006 &                &  1.220 ~~0.006 &  0.953 ~~0.006 \\
RS Oph               & 53997.6143 & 11.395 ~~0.004 &  1.595 ~~0.006 &                &  1.218 ~~0.004 &  0.956 ~~0.006 \\
RS Oph               & 53997.6204 & 11.406 ~~0.003 &  1.582 ~~0.006 &                &  1.206 ~~0.006 &  0.970 ~~0.006 \\
RS Oph               & 54003.6142 & 11.498 ~~0.005 &  1.531 ~~0.005 &                &  1.184 ~~0.006 &  1.022 ~~0.007 \\
RS Oph               & 54003.6221 & 11.484 ~~0.005 &  1.487 ~~0.005 &                &  1.189 ~~0.007 &  1.021 ~~0.007 \\
RS Oph               & 54010.5971 & 11.572 ~~0.004 &  1.349 ~~0.006 &                &  1.157 ~~0.006 &  1.080 ~~0.007 \\
RS Oph               & 54010.6049 & 11.586 ~~0.004 &  1.361 ~~0.006 &                &  1.180 ~~0.005 &  1.047 ~~0.005 \\
RS Oph               & 54022.6085 & 11.528 ~~0.004 &  1.380 ~~0.007 &                &  1.167 ~~0.007 &  1.095 ~~0.009 \\
RS Oph               & 54022.6167 & 11.611 ~~0.023 &  1.330 ~~0.024 &                &  1.256 ~~0.033 &  1.004 ~~0.035 \\
RT Ser               & 52157.6166 & 14.734 ~~0.005 &  1.367 ~~0.009 &  0.067 ~~0.012 &  1.890 ~~0.006 &  1.809 ~~0.003 \\
RT Ser               & 52171.6237 & 14.929 ~~0.004 &  1.249 ~~0.006 &--0.005 ~~0.009 &  1.902 ~~0.005 &  1.828 ~~0.004 \\
RT Ser               & 52174.6012 & 14.962 ~~0.005 &  1.232 ~~0.009 &--0.019 ~~0.011 &  1.916 ~~0.006 &  1.844 ~~0.004 \\
RW Hya               & 52337.9123 &  8.805 ~~0.002 &  1.380 ~~0.002 &--0.087 ~~0.003 &  0.931 ~~0.002 &  0.841 ~~0.003 \\
RW Hya               & 52382.8409 &  8.709 ~~0.002 &  1.401 ~~0.004 &  0.162 ~~0.004 &  0.883 ~~0.003 &  0.923 ~~0.002 \\
RW Hya               & 52383.8427 &  8.733 ~~0.002 &  1.419 ~~0.002 &  0.141 ~~0.001 &  0.917 ~~0.002 &  0.899 ~~0.002 \\
R Aqr                & 52186.7375 &  9.112 ~~0.003 &  1.334 ~~0.002 &--0.668 ~~0.006 &                &                \\
R Aqr                & 52288.5723 & 11.124 ~~0.001 &  1.173 ~~0.002 &--0.602 ~~0.004 &                &                \\
R Aqr                & 52292.5669 & 11.106 ~~0.003 &  1.177 ~~0.003 &--0.592 ~~0.002 &                &                \\
StH$\alpha$ 149      & 51694.8565 & 12.089 ~~0.002 &  1.538 ~~0.001 &  0.618 ~~0.003 &  1.117 ~~0.002 &  1.360 ~~0.001 \\
StH$\alpha$ 149      & 51695.8420 & 12.109 ~~0.001 &  1.542 ~~0.002 &  0.628 ~~0.004 &  1.140 ~~0.001 &  1.328 ~~0.002 \\
StH$\alpha$ 149      & 51728.8013 & 11.980 ~~0.001 &  1.576 ~~0.002 &  0.704 ~~0.004 &  1.122 ~~0.001 &  1.294 ~~0.002 \\
StH$\alpha$ 164      & 51728.8973 & 14.455 ~~0.002 &  2.047 ~~0.006 &  1.293 ~~0.032 &  1.490 ~~0.003 &  1.729 ~~0.003 \\
StH$\alpha$ 164      & 51729.8674 & 14.481 ~~0.003 &  2.026 ~~0.008 &  1.228 ~~0.028 &  1.488 ~~0.004 &  1.754 ~~0.003 \\
StH$\alpha$ 164      & 51730.8442 & 14.492 ~~0.003 &  2.016 ~~0.008 &  1.283 ~~0.027 &  1.508 ~~0.004 &  1.732 ~~0.002 \\
StH$\alpha$ 169      & 52110.9392 & 13.742 ~~0.002 &  1.638 ~~0.004 &  0.890 ~~0.008 &  1.056 ~~0.003 &  1.099 ~~0.003 \\
StH$\alpha$ 169      & 52145.8670 & 13.642 ~~0.002 &  1.634 ~~0.004 &  0.989 ~~0.009 &  1.026 ~~0.003 &  1.101 ~~0.003 \\
StH$\alpha$ 169      & 52146.8374 & 13.647 ~~0.002 &  1.645 ~~0.004 &  0.978 ~~0.008 &  1.032 ~~0.003 &  1.096 ~~0.003 \\
StH$\alpha$ 180      & 51728.9135 & 12.667 ~~0.001 &  1.409 ~~0.002 &  0.121 ~~0.004 &  0.921 ~~0.001 &  0.873 ~~0.002 \\
StH$\alpha$ 180      & 51729.8912 & 12.681 ~~0.001 &  1.398 ~~0.002 &  0.124 ~~0.004 &  0.923 ~~0.001 &  0.885 ~~0.002 \\
StH$\alpha$ 180      & 51730.9055 & 12.676 ~~0.001 &  1.400 ~~0.002 &  0.114 ~~0.004 &  0.923 ~~0.001 &  0.877 ~~0.002 \\
StH$\alpha$ 190      & 51729.9165 & 10.527 ~~0.002 &  0.836 ~~0.002 &--0.239 ~~0.002 &  0.496 ~~0.002 &  0.479 ~~0.002 \\
StH$\alpha$ 190      & 51730.9527 & 10.500 ~~0.001 &  0.840 ~~0.001 &--0.225 ~~0.002 &  0.491 ~~0.001 &  0.468 ~~0.002 \\
StH$\alpha$ 190      & 51731.8393 & 10.484 ~~0.001 &  0.837 ~~0.002 &--0.215 ~~0.002 &  0.498 ~~0.001 &  0.474 ~~0.002 \\
StH$\alpha$ 32       & 51515.7193 & 12.781 ~~0.002 &  1.462 ~~0.004 &--0.378 ~~0.005 &  0.778 ~~0.004 &  0.593 ~~0.004 \\
StH$\alpha$ 32       & 51528.7647 & 12.810 ~~0.001 &  1.437 ~~0.002 &--0.389 ~~0.003 &  0.787 ~~0.002 &  0.599 ~~0.005 \\
StH$\alpha$ 32       & 51547.6590 & 12.706 ~~0.001 &  1.448 ~~0.001 &--0.240 ~~0.003 &  0.757 ~~0.001 &  0.605 ~~0.002 \\
StH$\alpha$ 32       & 51548.6358 & 12.708 ~~0.001 &  1.453 ~~0.002 &--0.222 ~~0.003 &  0.765 ~~0.002 &  0.589 ~~0.003 \\
StH$\alpha$ 32       & 51549.6589 & 12.706 ~~0.001 &  1.455 ~~0.002 &--0.212 ~~0.003 &  0.763 ~~0.001 &  0.600 ~~0.001 \\
TX CVn               & 51614.8067 & 10.092 ~~0.003 &  0.691 ~~0.002 &  0.130 ~~0.004 &  0.647 ~~0.003 &  0.748 ~~0.002 \\
TX CVn               & 51629.8527 & 10.040 ~~0.002 &  0.708 ~~0.001 &  0.117 ~~0.007 &  0.629 ~~0.004 &  0.765 ~~0.003 \\
TX CVn               & 51639.8016 & 10.170 ~~0.002 &  0.659 ~~0.002 &  0.069 ~~0.007 &  0.561 ~~0.003 &  0.917 ~~0.003 \\
T CrB                & 52382.8510 & 10.073 ~~0.001 &  1.402 ~~0.001 &  0.536 ~~0.003 &  1.088 ~~0.002 &  1.458 ~~0.001 \\
T CrB                & 52383.8649 & 10.091 ~~0.001 &  1.406 ~~0.002 &  0.535 ~~0.004 &  1.080 ~~0.002 &  1.495 ~~0.004 \\
T CrB                & 52385.9104 & 10.177 ~~0.003 &  1.423 ~~0.002 &  0.658 ~~0.002 &  1.102 ~~0.001 &  1.515 ~~0.003 \\
\tablerule
\end{tabular}
}
\end{center}
\vspace{-2mm}
\begin{center}
\vbox{\footnotesize
\tabcolsep=15pt
\begin{tabular}{@{~~}l@{~~}c@{~~}r@{~~}r@{~~}r@{~~}r@{~~}r@{~~}}
\multicolumn{7}{c}{\parbox{90mm}{
{\normbf \ \ Table 1.}{\norm\ (continues).}}}\\
\tablerule
\multicolumn{1}{c}{name}&
\multicolumn{1}{c}{HJD}&
\multicolumn{1}{c}{V}&
\multicolumn{1}{c}{B-V}&
\multicolumn{1}{c}{U-B}&
\multicolumn{1}{c}{V-R$_c$}&
\multicolumn{1}{c}{R$_c$-I$_c$}\\
\tablerule
UKS Ce1              & 51664.8863 & 15.639 ~~0.005 &  1.460 ~~0.011 &                &                &                \\
UKS Ce1              & 51692.8099 & 15.872 ~~0.010 &  1.854 ~~0.021 &                &  1.006 ~~0.013 &  0.834 ~~0.011 \\
UKS Ce1              & 51693.8293 & 15.886 ~~0.011 &  1.839 ~~0.022 &  1.127 ~~0.094 &  1.047 ~~0.014 &  0.800 ~~0.012 \\
UU Ser               & 51692.8722 & 15.474 ~~0.006 &  1.776 ~~0.013 &  2.456 ~~0.128 &  0.980 ~~0.008 &  0.890 ~~0.007 \\
UU Ser               & 51693.8563 & 15.474 ~~0.006 &  1.778 ~~0.014 &  2.156 ~~0.092 &  0.984 ~~0.008 &  0.890 ~~0.007 \\
UU Ser               & 51695.8653 & 15.461 ~~0.005 &  1.813 ~~0.012 &  2.323 ~~0.105 &  0.969 ~~0.007 &  0.891 ~~0.007 \\
UV Aur               & 52186.9209 &  9.508 ~~0.007 &  3.890 ~~0.046 &  2.131 ~~0.117 &                &                \\
UV Aur               & 52287.7649 & 11.054 ~~0.011 &  5.165 ~~0.065 &  1.752 ~~0.258 &                &                \\
UV Aur               & 52288.7382 & 11.064 ~~0.015 &  4.908 ~~0.157 &                &                &                \\
V1016 Cyg            & 52145.8875 & 11.261 ~~0.001 &  0.426 ~~0.002 &--0.901 ~~0.002 &  1.031 ~~0.001 &--0.101 ~~0.002 \\
V1016 Cyg            & 52146.8650 & 11.263 ~~0.001 &  0.414 ~~0.002 &--0.893 ~~0.001 &  1.039 ~~0.001 &--0.121 ~~0.002 \\
V1016 Cyg            & 52153.8122 & 11.260 ~~0.002 &  0.425 ~~0.002 &--0.882 ~~0.002 &  1.041 ~~0.002 &--0.083 ~~0.002 \\
V1329 Cyg            & 51873.6384 & 13.193 ~~0.002 &  0.681 ~~0.003 &--0.770 ~~0.003 &  1.376 ~~0.002 &  1.097 ~~0.002 \\
V1329 Cyg            & 51877.6264 & 13.195 ~~0.002 &  0.664 ~~0.003 &--0.788 ~~0.003 &  1.363 ~~0.003 &  1.066 ~~0.002 \\
V1329 Cyg            & 51906.5564 & 13.102 ~~0.002 &  0.694 ~~0.004 &--0.929 ~~0.004 &  1.388 ~~0.003 &  1.144 ~~0.002 \\
V1413 Aql            & 51453.7053 & 13.222 ~~0.003 &  0.945 ~~0.003 &  0.044 ~~0.004 &  0.929 ~~0.003 &  1.149 ~~0.003 \\
V1413 Aql            & 51457.7294 & 13.073 ~~0.003 &  0.960 ~~0.003 &  0.040 ~~0.004 &  0.928 ~~0.004 &  1.094 ~~0.005 \\
V1413 Aql            & 51464.7005 & 12.987 ~~0.004 &  0.979 ~~0.004 &  0.038 ~~0.003 &  0.918 ~~0.005 &  1.099 ~~0.005 \\
V335 Vul             & 51349.9180 & 12.901 ~~0.003 &  5.125 ~~0.017 &                &  1.996 ~~0.002 &  1.460 ~~0.001 \\
V335 Vul             & 51352.8235 & 12.865 ~~0.003 &  5.096 ~~0.026 &                &  2.007 ~~0.002 &  1.450 ~~0.001 \\
V335 Vul             & 51398.8497 & 11.976 ~~0.001 &  4.671 ~~0.006 &                &  1.669 ~~0.002 &  1.283 ~~0.002 \\
V335 Vul             & 51402.8674 & 11.957 ~~0.001 &  4.602 ~~0.005 &                &  1.641 ~~0.001 &  1.122 ~~0.002 \\
V335 Vul             & 51515.5501 & 11.310 ~~0.004 &  3.055 ~~0.005 &                &  1.452 ~~0.001 &  1.276 ~~0.001 \\
V352 Aql             & 52153.7688 & 16.696 ~~0.023 &  1.678 ~~0.077 &  0.352 ~~0.216 &  1.795 ~~0.025 &  1.739 ~~0.012 \\
V352 Aql             & 52171.7364 & 16.555 ~~0.011 &  1.776 ~~0.027 &  0.540 ~~0.055 &  1.729 ~~0.013 &  1.714 ~~0.007 \\
V352 Aql             & 52174.7022 & 16.569 ~~0.011 &  1.763 ~~0.027 &  0.481 ~~0.053 &  1.750 ~~0.013 &  1.714 ~~0.007 \\
V4018 Sgr            & 52051.9292 & 13.144 ~~0.004 &  1.086 ~~0.004 &                &                &                \\
V4018 Sgr            & 52146.6819 & 13.359 ~~0.003 &  1.087 ~~0.005 &--0.441 ~~0.005 &  1.147 ~~0.004 &  1.483 ~~0.003 \\
V4018 Sgr            & 52153.6860 & 13.336 ~~0.004 &  1.087 ~~0.007 &--0.383 ~~0.009 &  1.187 ~~0.005 &  1.397 ~~0.004 \\
V4018 Sgr            & 52157.6502 & 13.318 ~~0.003 &  1.101 ~~0.006 &--0.393 ~~0.007 &  1.169 ~~0.004 &  1.401 ~~0.003 \\
V407 Cyg             & 51316.9262 & 11.755 ~~0.001 &  1.522 ~~0.001 &  0.112 ~~0.002 &  1.934 ~~0.002 &  2.193 ~~0.002 \\
V407 Cyg             & 51317.9619 & 11.685 ~~0.001 &  1.482 ~~0.001 &  0.012 ~~0.002 &  1.845 ~~0.001 &  1.801 ~~0.002 \\
V407 Cyg             & 51318.9649 & 11.570 ~~0.001 &  1.456 ~~0.001 &  0.052 ~~0.002 &  1.804 ~~0.001 &  2.204 ~~0.003 \\
V4074 Sgr            & 52382.9889 & 13.802 ~~0.051 &                &                &  1.058 ~~0.057 &  1.545 ~~0.033 \\
V4074 Sgr            & 52383.9625 & 13.871 ~~0.047 &                &                &  1.082 ~~0.055 &  1.531 ~~0.041 \\
V4074 Sgr            & 52385.9939 & 13.684 ~~0.045 &                &                &  1.032 ~~0.059 &  1.423 ~~0.046 \\
V4368 Sgr            & 51728.8271 & 10.536 ~~0.002 &  0.657 ~~0.002 &  0.019 ~~0.001 &  0.373 ~~0.003 &  0.446 ~~0.002 \\
V4368 Sgr            & 51729.8203 & 10.616 ~~0.002 &  0.619 ~~0.001 &  0.019 ~~0.002 &  0.421 ~~0.004 &  0.414 ~~0.004 \\
V4368 Sgr            & 51730.8205 & 10.574 ~~0.001 &  0.626 ~~0.001 &  0.048 ~~0.001 &  0.383 ~~0.002 &  0.433 ~~0.002 \\
V443 Her             & 51315.9769 & 11.362 ~~0.002 &  1.061 ~~0.003 &--0.420 ~~0.003 &  1.302 ~~0.003 &  1.525 ~~0.002 \\
V443 Her             & 51316.9094 & 11.369 ~~0.002 &  1.054 ~~0.003 &--0.428 ~~0.003 &  1.301 ~~0.003 &  1.507 ~~0.001 \\
V443 Her             & 51318.8944 & 11.386 ~~0.002 &  1.054 ~~0.003 &--0.422 ~~0.004 &  1.295 ~~0.002 &  1.537 ~~0.002 \\
V503 Her             & 51694.8318 & 12.692 ~~0.002 &  1.216 ~~0.002 &  0.851 ~~0.004 &  0.704 ~~0.002 &  0.716 ~~0.001 \\
V503 Her             & 51695.8099 & 12.691 ~~0.001 &  1.222 ~~0.002 &  0.857 ~~0.004 &  0.702 ~~0.001 &  0.714 ~~0.001 \\
V503 Her             & 51728.7907 & 12.635 ~~0.001 &  1.391 ~~0.002 &  1.211 ~~0.004 &  0.781 ~~0.001 &  0.721 ~~0.001 \\
V627 Cas             & 51402.9163 & 12.882 ~~0.001 &  2.595 ~~0.003 &  1.159 ~~0.008 &  2.132 ~~0.001 &  2.098 ~~0.002 \\
V627 Cas             & 51403.8571 & 12.863 ~~0.001 &  2.604 ~~0.003 &  1.173 ~~0.007 &  2.123 ~~0.001 &  2.114 ~~0.002 \\
V627 Cas             & 51453.7941 & 12.535 ~~0.001 &  2.680 ~~0.003 &  1.071 ~~0.007 &  2.052 ~~0.001 &  2.012 ~~0.003 \\
V694 Mon             & 51614.7186 & 10.876 ~~0.001 &  0.425 ~~0.002 &--0.325 ~~0.005 &  0.776 ~~0.002 &  1.488 ~~0.001 \\
V694 Mon             & 51629.7111 & 10.877 ~~0.001 &  0.490 ~~0.002 &--0.400 ~~0.003 &  0.807 ~~0.002 &  1.483 ~~0.002 \\
V694 Mon             & 51639.6293 & 10.676 ~~0.001 &  0.436 ~~0.002 &--0.363 ~~0.002 &  0.713 ~~0.001 &  1.409 ~~0.002 \\
V471 Per             & 51515.6763 & 13.088 ~~0.001 &  0.986 ~~0.001 &  0.041 ~~0.002 &  0.762 ~~0.002 &  0.385 ~~0.004 \\
V471 Per             & 51549.7426 & 13.093 ~~0.001 &  0.999 ~~0.002 &  0.040 ~~0.004 &  0.763 ~~0.002 &  0.373 ~~0.004 \\
V471 Per             & 51550.7318 & 13.099 ~~0.002 &  0.997 ~~0.003 &  0.015 ~~0.003 &  0.767 ~~0.004 &  0.373 ~~0.004 \\
\tablerule
\end{tabular}
}
\end{center}
\vspace{-2mm}
\begin{center}
\vbox{\footnotesize
\tabcolsep=15pt
\begin{tabular}{@{~~}l@{~~}c@{~~}r@{~~}r@{~~}r@{~~}r@{~~}r@{~~}}
\multicolumn{7}{c}{\parbox{90mm}{
{\normbf \ \ Table 1.}{\norm\ (continues).}}}\\
\tablerule
\multicolumn{1}{c}{name}&
\multicolumn{1}{c}{HJD}&
\multicolumn{1}{c}{V}&
\multicolumn{1}{c}{B-V}&
\multicolumn{1}{c}{U-B}&
\multicolumn{1}{c}{V-R$_c$}&
\multicolumn{1}{c}{R$_c$-I$_c$}\\
\tablerule
V919 Sgr             & 51456.6430 & 13.023 ~~0.003 &  1.196 ~~0.004 &--0.328 ~~0.003 &  1.302 ~~0.003 &  1.462 ~~0.002 \\
V919 Sgr             & 51457.6074 & 13.012 ~~0.003 &  1.190 ~~0.002 &--0.350 ~~0.005 &  1.291 ~~0.001 &  1.472 ~~0.001 \\
V919 Sgr             & 51465.6320 & 12.928 ~~0.004 &  1.221 ~~0.004 &--0.260 ~~0.004 &  1.285 ~~0.004 &  1.441 ~~0.003 \\
V934 Her             & 52382.8879 &  7.635 ~~0.002 &  1.606 ~~0.001 &  1.837 ~~0.005 &  0.896 ~~0.003 &  1.044 ~~0.002 \\
V934 Her             & 52383.8863 &  7.633 ~~0.002 &  1.611 ~~0.001 &  1.914 ~~0.003 &  0.896 ~~0.002 &  1.052 ~~0.001 \\
V934 Her             & 52385.9454 &  7.643 ~~0.003 &  1.614 ~~0.001 &  1.900 ~~0.001 &  0.956 ~~0.002 &  1.122 ~~0.001 \\
Wray 15-1470         & 52052.8286 & 13.408 ~~0.002 &  0.745 ~~0.004 &--0.630 ~~0.006 &  1.520 ~~0.003 &  0.947 ~~0.001 \\
Wray 15-1470         & 52061.8293 & 13.464 ~~0.003 &  0.722 ~~0.005 &--0.664 ~~0.008 &  1.529 ~~0.002 &  0.960 ~~0.002 \\
Wray 15-1470         & 52115.6868 & 13.215 ~~0.002 &  0.853 ~~0.004 &--0.564 ~~0.006 &  1.472 ~~0.004 &  0.958 ~~0.004 \\
Wray 15-157          & 51999.6359 & 13.400 ~~0.003 &  1.440 ~~0.007 &  0.686 ~~0.012 &  0.856 ~~0.005 &  0.747 ~~0.005 \\
Wray 15-157          & 52014.6136 & 13.552 ~~0.007 &  1.471 ~~0.020 &  0.675 ~~0.044 &  0.870 ~~0.009 &  0.763 ~~0.008 \\
Wray 15-157          & 52016.6184 & 13.569 ~~0.007 &  1.465 ~~0.017 &  0.712 ~~0.033 &  0.867 ~~0.009 &  0.750 ~~0.008 \\
YY Her               & 51729.7533 & 12.934 ~~0.001 &  1.373 ~~0.002 &  0.013 ~~0.004 &  1.073 ~~0.001 &  1.293 ~~0.001 \\
YY Her               & 51730.7664 & 12.922 ~~0.001 &  1.373 ~~0.002 &  0.024 ~~0.004 &  1.088 ~~0.001 &  1.252 ~~0.001 \\
YY Her               & 51731.7535 & 12.919 ~~0.001 &  1.380 ~~0.002 &  0.029 ~~0.004 &  1.087 ~~0.001 &  1.259 ~~0.001 \\
ZZ CMi               & 52186.9821 &  9.974 ~~0.001 &  1.457 ~~0.001 &  0.626 ~~0.003 &  1.628 ~~0.002 &                \\
ZZ CMi               & 52287.7843 &  9.826 ~~0.000 &  1.336 ~~0.001 &  0.960 ~~0.001 &  1.410 ~~0.000 &  1.629 ~~0.000 \\
ZZ CMi               & 52288.7767 &  9.742 ~~0.000 &  1.412 ~~0.001 &  1.009 ~~0.001 &  1.405 ~~0.000 &                \\
Z And                & 52186.7214 &  9.666 ~~0.002 &  0.675 ~~0.001 &--0.817 ~~0.003 &  0.892 ~~0.002 &  1.040 ~~0.002 \\
Z And                & 52288.6288 &  9.960 ~~0.002 &  0.680 ~~0.002 &--0.800 ~~0.002 &  0.868 ~~0.001 &  1.013 ~~0.001 \\
Z And                & 52292.6282 & 10.026 ~~0.000 &  0.719 ~~0.001 &--0.799 ~~0.002 &  0.994 ~~0.000 &  1.126 ~~0.002 \\
Z And                & 52566.6292 & 10.066 ~~0.003 &  0.892 ~~0.004 &--0.549 ~~0.005 &                &                \\
Z And                & 53251.6659 &  9.193 ~~0.004 &  0.620 ~~0.006 &--0.394 ~~0.004 &  0.767 ~~0.002 &  0.918 ~~0.003 \\
Z And                & 53293.8157 &  9.399 ~~0.005 &  0.655 ~~0.003 &                &  0.804 ~~0.011 &  1.015 ~~0.007 \\
Z And                & 54476.6710 &  9.355 ~~0.002 &  0.608 ~~0.004 &                &  0.700 ~~0.004 &  0.801 ~~0.005 \\
Z And                & 54476.6731 &  9.362 ~~0.004 &  0.604 ~~0.003 &                &  0.660 ~~0.003 &  0.711 ~~0.005 \\
Z And                & 54477.5773 &  9.399 ~~0.003 &  0.705 ~~0.004 &                &  0.800 ~~0.005 &  0.688 ~~0.003 \\
Z And                & 54477.5795 &  9.448 ~~0.003 &  0.665 ~~0.005 &                &  0.802 ~~0.004 &  0.730 ~~0.003 \\
Z And                & 54479.5757 &  9.501 ~~0.004 &  0.683 ~~0.006 &                &  0.912 ~~0.005 &  0.751 ~~0.003 \\
Z And                & 54479.5778 &  9.501 ~~0.003 &  0.686 ~~0.006 &                &  0.900 ~~0.004 &                \\
Z And                & 54480.6245 &  9.628 ~~0.004 &  0.672 ~~0.005 &                &  0.971 ~~0.006 &  0.899 ~~0.004 \\
Z And                & 54480.6265 &  9.628 ~~0.003 &  0.666 ~~0.005 &                &  0.986 ~~0.005 &  0.883 ~~0.003 \\
Z And                & 54484.5734 &  9.612 ~~0.006 &  0.666 ~~0.005 &                &  0.806 ~~0.005 &  0.721 ~~0.004 \\
Z And                & 54484.5762 &  9.662 ~~0.005 &  0.616 ~~0.005 &                &  0.774 ~~0.006 &  0.839 ~~0.006 \\
\tablerule
\end{tabular}
}
\end{center}
\vspace{-2mm}

\begin{center}
\vbox{\footnotesize
\tabcolsep=15pt
\begin{tabular}{cr}
\multicolumn{2}{c}{\parbox{48mm}{
{\normbf \ \ Table 2.}{\norm\ Data for V1261 Ori.}}}\\
\tablerule
52186.9578 & B=~~8.607  ~~0.001  \\     
52186.9604 & U=10.250  ~~0.003  \\   
52287.7433 & B=~~8.836  ~~0.002  \\   
52287.7454 & U=10.095  ~~0.004  \\   
52288.7302 & B=~~8.707  ~~0.002  \\   
52288.7324 & U=10.116  ~~0.003  \\   
\tablerule
\end{tabular}
}
\end{center}
\vspace{-2mm}

\References

 \refb  Belczynski~K., Mikolajewska~J., Munari~U., Ivison~R.~J., Friedjung~M. 2000, A\&AS 146, 407
 \refb  Henden~A., Munari~U. 2000, A\&AS 143, 343
 \refb  Henden~A., Munari~U. 2001, A\&A 372, 145
 \refb  Henden~A., Munari~U. 2006, A\&A 458, 339
 \refb  Honeycutt~R.~K. 1992, PASP 104, 435
 \refb  Landolt~A.~U. 1983, AJ 88, 439
 \refb  Landolt~A.~U. 1992, AJ 104, 340
 \refb  Munari~U., Yudin~B.~F., Taranova~O.~G., Massone~G., Marang~F., Roberts~G., Winkler~H., Whitelock~P.~A. 1992, A\&AS 93, 383

\end{document}